\documentclass[prx,showpacs,twocolumn,amsmath,amssymb,floatfix]{revtex4-1}

\usepackage{hyperref}
\usepackage{graphicx}
\usepackage{bm}
\usepackage{color}
\usepackage[applemac]{inputenc}
\usepackage{amssymb}

\begin{document}  
   
\title{Quantifying wave-function overlaps in inhomogeneous Majorana nanowires}
\author{Fernando Pe\~naranda$^1$, Ram\'on Aguado$^2$, Pablo San-Jose$^2$, Elsa Prada$^1$}
\affiliation{$^1$Departamento de F\'isica de la Materia Condensada, Condensed Matter Physics Center (IFIMAC) and Instituto Nicol\'as Cabrera, Universidad Aut\'onoma de Madrid, E-28049 Madrid, Spain\\
$^2$Materials Science Factory, Instituto de Ciencia de Materiales de Madrid (ICMM), Consejo Superior de Investigaciones Cient\'{i}ficas (CSIC), Sor Juana In\'{e}s de la Cruz 3, 28049 Madrid, Spain}

\date{\today} 

\begin{abstract}
A key property of Majorana zero modes is their protection against local perturbations. 
{In the standard picture, this protection is {guaranteed by a} high degree of spatial nonlocality of {the} Majoranas, {namely a suppressed} wave-function overlap, {in the topological phase.}}
{However}, a careful {characterisation of resilience to local noise} goes beyond {mere} spatial separation, and must also take into account the projection of wave-function spin. 
By considering the susceptibility of a given zero mode to different local perturbations, {we find the relevant forms of spin-resolved wave-function overlaps that measure its resilience.}
We quantify {these} 
{overlaps} and study {their} dependence with nanowire parameters in several classes of experimentally relevant configurations. {These include nanowires with inhomogeneous depletion and induced pairing, barriers and quantum dots. Smooth inhomogeneities have been shown to produce near-zero modes, so-called pseudo-Majoranas, below the critical Zeeman field in the bulk. Surprisingly, their resilience is found to be comparable or better than that of topological Majoranas in realistic systems.}  
We {further} study how accurately their {overlaps} can be estimated using a purely local measurement on one end of the nanowire, accessible through conventional transport experiments. 
In uniform nanowires this local estimator is remarkably accurate. In inhomogeneous cases it is less accurate but can still provide reasonable estimates for potential inhomogeneities of the order of the superconducting gap. We further analyse the {zero mode} wave-function structure, spin texture and spectral features associated with each type of inhomogeneity. All our results highlight the strong connection between internal wave-function degrees of freedom, nonlocality and protection in smoothly inhomogeneous nanowires.
\end{abstract}

\maketitle

\section{Introduction}

A unique electronic state by the name of Majorana zero mode (MZM)\cite{Kitaev:PU01} associated with topological superconductivity has been the subject of intense research recently. The pace picked up after the first experimental hints of its existence were reported six years ago~\cite{Mourik:S12} in so-called Majorana nanowires, i.e. semiconducting nanowires with induced superconductivity and spin-orbit coupling subjected to a Zeeman field above a critical value $B>B_c$. These pioneering experiments were quickly followed by others~\cite{Deng:NL12,Das:NP12,Churchill:PRB13,Lee:NN14,Deng:S16,Zhang:N18,Grivnin:A18}, mostly revolving around robust zero energy midgap states in tunneling spectroscopy.
\begin{figure}
   \centering
   \includegraphics[width=\columnwidth]{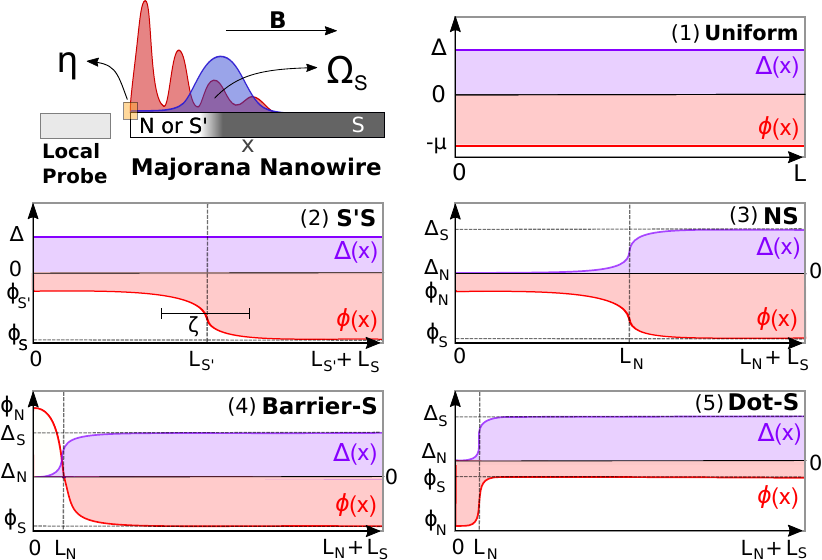}
   \caption{\textbf{Inhomogeneous nanowires.} Sketch of an inhomogeneous nanowire, hosting Majorana zero modes of overlap $\Omega_s$. The overlap may be estimated by a local quantity $\eta$ measured by a local probe. Five types of inhomogeneous profiles of the electrostatic potential $\phi(x)$ and pairing $\Delta(x)$ are considered: uniform,  {S'S (superconductor-superconductor), NS (normal-superconductor)}, Barrier-S and Dot-S. The latter two are subtypes of the general NS case. }
   \label{fig:sketch}
\end{figure}
From a technological perspective, MZMs are viewed by many as a possible foundation of a new form of quantum computer {platform} that could achieve topologically protection against some forms of {logic errors~\cite{Kitaev:PU01,Nayak:RMP08,Cheng:PRB12a,Sarma:NQI15} using a variety of architectures~\cite{Plugge:NJP17,Karzig:PRB17,Aasen:PRX16, Schrade:PRL18}}.
 
{A MZM is defined as a self-conjugate zero energy eigenstate inside a superconducting gap and it is localised in space to some degree{~\cite{Kitaev:PU01, Oreg:PRL10, Lutchyn:PRL10}}. The most common place to find a MZM is at boundaries between regions of distinct electronic topology~\cite{Qi:RMP11,Aguado:RNC17,Lutchyn:NRM18}. From the point of view of fundamental physics, standard theory predicts that a MZM should exhibit some truly exotic properties. Its second-quantised operator $\gamma$ satisfies the Majorana relation $\gamma^\dagger = \gamma$, i.e. a MZM quasiparticle is its own ``antiparticle''.} A MZM behaves in many respects as half an electron, with each MZM emerging simultaneously with a second Majorana partner located at some other position in the system. Two such ``electron-halves'' form a rather unusual, spatially nonlocal fermion{~\cite{Kitaev:PU01,Jackiw:PS12,Fu:PRL10,Semenoff:C16}}. The nonlocal nature of this fermion pins it to zero energy regardless of any local perturbations performed on either of the MZMs. This is often called  topological protection{~\cite{Kitaev:PU01,Cheng:PRB12a}}, although protection through nonlocality is perhaps a better description, as will be argued here. Each MZM also exhibits non-Abelian braiding statistics upon exchange~\cite{Nayak:RMP08}, {and obey the same fusion rules as fractionalised non-Abelian Ising anyons}{~\cite{Kitaev:AOP06,Bonderson:PRB13,Barkeshli:A14}}. All these properties are expected to be remarkably robust, and to not require any fine tuning of the system's state. The reason is that, at least within standard theory, they are a consequence of the different band topology at either side of the boundary they inhabit, which does not depend on microscopic details. 

{Strictly speaking, however, the topological protection of MZMs is only exact for boundaries between semi-infinite systems, and becomes approximate for finite-sized systems {where the overlap between Majorana wavefunctions is finite.}   
Topological band theory also fails to account} for the properties {and potential protection} of so-called trivial Andreev zero modes, also known as pseudo-MZMs or quasi-MZMs, predicted to appear in smoothly inhomogeneous nanowires without any obvious form of band-topological order~\cite{Kells:PRB12,Prada:PRB12,Liu:PRB17,Moore:PRB18,Setiawan:PRB17,Moore:PRB18a,Liu:PRB18,Vuik:18}. An example of such states relevant to the present work arises in fully trivial nanowires ($B<B_c$) hosting a sufficiently smooth normal-superconductor interface, wherein modes of arbitrarily small energy localise. 

All the experimental evidence so far of conventional topological MZMs can be mimicked by pseudo-MZMs. This realisation has given rise to an intense debate regarding possible loopholes in the interpretation of the experimental observations, and on the protection, or lack thereof, of the observed zero modes.  
Intriguingly, these states share most properties with MZMs at the end of a uniform $B>B_c$ topological nanowire, except in one crucial aspect: {their wavefunctions are not concentrated at opposite ends of the nanowire, so their overlap is not controlled by extrinsic device parameters like nanowire length $L$. Since spatial nonlocality is necessary to achieve resilience of MZMs against generic error-inducing, parity-conserving} local perturbations, it is often argued that pseudo-MZMs, unlike MZMs, would not be useful for topological quantum computation.

{We here reassess this assumption by comparing the overlaps of topological-MZMs and pseudo-MZMs in realistic finite nanowires. In such systems topological MZMs can exhibit sizable overlaps, not necessarily smaller than those of pseudo-MZMs. Despite the fact that topological-MZM overlaps are ideally an exponentially decreasing function of length $L$, $\Omega\sim e^{-L/\xi}$, the coherence length $\xi$ is not necessarily small, and actually tends to grow with magnetic field. This leads to an expected increase of overlaps as one enters deep into the topological phase~\cite{Prada:PRB12,Das-Sarma:PRB12,Mishmash:PRB16,Dominguez:NQM17}. In contrast, we find that pseudo-MZM from sufficiently smooth potentials can develop wavefunctions with small overlaps very quickly (they have a different --Gaussian-- profile than topological MZMs). Thus, there is no fundamental reason that dictates which of the two types of MZMs, topological or non-topological, is likely to be better protected against local perturbations in realistic, micron-length nanowire devices}. 

{Motivated by this, we here characterise MZMs purely in terms of their wave-function overlaps instead of classifying states into trivial and non-trivial based on bulk topological invariants. Specifically, we will focus on different measures of Majorana wave-function nonlocality, which quantify the susceptibility to arbitrary local perturbations that preserve fermion parity. Formally, the associated susceptibilities will be} expressed as different spatial overlap integrals $\Omega$ of the Majorana Nambu-spinorial wavefunctions, depending on the type of perturbation. Despite all of these integrals expressing nonlocality, the way the internal spin degrees of freedom combine in the overlap integral is different. This leads to several definitions of {the degree of} nonlocality $0\leq 1-\Omega\leq 1$ that go beyond purely spatial separation, and that are directly connected to protection of MZMs {against different, parity-preserving, local} perturbations~\cite{Kitaev:PU01,Cheng:PRB12a}. The quantity  $1-\Omega$ thus takes the meaning of a figure-of-merit of a pair of zero modes, irrespective of their topological {origin}, which is applicable in {isolated systems of arbitrary length, where the distinction between pseudo-MZMs and `proper' topological MZMs is ill-defined}.

As an aside, we note that an alternative theoretical framework has been recently proposed that allows to recover a well-defined and unambiguous trivial/non-trivial classification within this continuum of MZMs of isolated systems. It defines the topological nature of these zero modes in more general terms by considering the exceptional-point topology of the non-hermitian Hamiltonian that describes the system when it is coupled to a reservoir~\cite{Avila:A18,Pikulin:JL12,Pikulin:PRB13}. In essence, the coupling to the reservoir makes the system infinite, so that it is once more amenable to a rigorous topological classification. This approach is related to band topology, but is more general, and in it the degree of nonlocality of the isolated states studied here plays a crucial role.

In this work we further consider the practical problem of quantifying and detecting the degree of nonlocality of a given zero mode using purely local measurements by local spectroscopic probes.  These include e.g. a tunnel contact or a quantum dot coupled to a certain spot in a Majorana nanowire~\cite{Flensberg:PRL11, Leijnse:PRB11, Leijnse:PRB12, Gharavi:PRB16, Prada:PRB17, Malciu:PRB18}, a setup routinely used today in the lab to perform tunnelling spectroscopy~\cite{Lee:NN14, Deng:S16, Vaitiekenas:PRL18, Deng:PRB18}, see Fig. \ref{fig:sketch}. This challenge seems a priori hopeless since quantifying nonlocality involves knowing the distribution of the zero mode throughout an extended region in space, not just at one spot. {Thus, it seems necessary to resort to complex nonlocal cross-correlation or interferometry detection schemes~\cite{Hell:PRB18,Liu:PRB13,Zocher:PRL13,Li:SR14,Zocher:PRL13,Haim:PRL15,Haim:PRB15, Moore:PRB18}}. We show, however, that the spatial distribution of subgap states is not completely arbitrary in realistic systems, but spans a finite volume in the space of all possible wavefunctions. {Due to this constraint, local measurements at one end of the nanowire} remain highly correlated with the actual Majorana wave-function overlap throughout the system.

This work is organised as follows. Section \ref{sec:nonlocality} presents the basic concepts and definitions of overlaps and local estimators, and the five types of nanowire configurations to be studied, see Fig. \ref{fig:sketch}. Section \ref{sec:uniform} is devoted to uniform nanowires. The basic Lutchyn-Oreg model is presented, together with its phenomenology regarding spectrum, zero-mode overlaps and their correlation with local estimators. Sections \ref{sec:S'S} and \ref{sec:NS} present the corresponding analysis in inhomogeneous superconductor-superconductor and normal-superconductor nanowires, respectively. In the latter case we also analyse specific barrier-superconductor and quantum dot-superconductor configurations, of relevance to many experimental devices. We finally present, in Sec. \ref{sec:spin}, a discussion of the spatial spin density of the Majorana wavefunction in the various types of inhomogeneous nanowires as a function of their smoothness. This will allow us to distinguish between two characteristic types of wavefunctions, that of conventional abrupt Majoranas at sharp insulating interfaces, and that of Gaussian-like smooth Majoranas that develop at smooth superconductor interfaces. Finally, in Sec. \ref{sec:conclusion} we conclude.

\section{nonlocality and local estimator}
\label{sec:nonlocality}

\subsection{Majorana basis}
Consider a generic subgap eigenstate $c$ in a quasi-1D superconductor 
\[
c = \int dx \sum_\sigma u_\sigma(x)\psi_\sigma(x) + v_\sigma(x)\psi^\dagger_\sigma(x).
\]
Here $\sigma$ denotes spin projections on a given axis, chosen in this work as the $x$ axis along which a Zeeman field will later be applied. This fermionic state can be decomposed into two Majorana components $\gamma_{L}$ and $\gamma_{R}$
\begin{eqnarray}
c &=& \frac{\gamma_L+i\gamma_R}{\sqrt{2}},\nonumber\\
c^\dagger &=& \frac{\gamma_L-i\gamma_R}{\sqrt{2}},
\label{decomposition}
\end{eqnarray}
so that 
\begin{equation}
\gamma_{L,R} = \int dx \sum_\sigma u^{L,R}_\sigma(x)\psi_\sigma(x) + \left[u^{L,R}_\sigma(x)\right]^*\psi^\dagger_\sigma(x).\nonumber
\end{equation}
By definition, $\gamma_{L,R}$ are self-conjugate, $\gamma_{L,R}=\gamma^\dagger_{L,R}$ (Majorana reality condition). Their wavefunctions $u^{L,R}_\sigma(x)$ can be expressed in terms of the particle and hole wave-function components $u_\sigma(x), v_\sigma(x)$ of eigenstate $c$ as
\begin{eqnarray}
u^{L}_\sigma(x) &=& \frac{u_\sigma(x) + v^*_\sigma(x)}{\sqrt{2}}, \nonumber\\
u^{R}_\sigma(x) &=& \frac{u_\sigma(x) - v^*_\sigma(x)}{i\sqrt{2}} .
\label{mwf}
\end{eqnarray}
All wavefunctions $u^{L,R}_\sigma(x), u_\sigma(x), v_\sigma(x)$ are normalised. In particular
\begin{equation}
\int dx \|\mathbf{u}^{L,R}(x)\|^2=1.
\label{normalisation}
\end{equation}
where $\mathbf{u}^{L,R}$ denotes the spinor of $u_\sigma^{L,R}$ components, and the $\|\dots\|$ denotes its norm.

Note that the Majorana decomposition of Eqs. (\ref{decomposition}, \ref{mwf}) is possible for \emph{any} Andreev state with finite energy $E_0$, not only for those with zero energy. Only if the subgap eigenstate $c$ has zero energy, the $\gamma_{L,R}$ will themselves be zero energy eigenstates. This is true \emph{regardless} of their spatial nonlocality or their topological/trivial origin. In this work we will always call these self-conjugate $\gamma_{L,R}$ zero modes, \emph{Majorana} zero modes (MZMs), since they satisfy the Majorana reality condition. We thus refer to MZMs independently of whether the system has a trivial or non-trivial band topology.

\subsection{Protection and wave-function nonlocality}

The standard definition of topological protection \cite{Sarma:NQI15} relies on Majorana midgap states with a sufficiently large gap to higher excitations and an exponentially suppressed energy, $E_0\sim e^{-L/\xi}$, resulting from spatial separation of Majorana wavefunctions, with $L$ the wire's length and $\xi$ the Majorana coherence length~\cite{Ben-Shach:PRB15}. This allows in practice to achieve a degenerate ground state that does not split in response to local perturbations. The spatial nonlocality of the Majoranas leads also to an exponentially suppressed sensitivity to arbitrary local perturbations. This is ultimately the reason why nonlocality is such a key Majorana property.

As argued in the introduction, this picture has to be extended in more general situations, wherein zero modes arise due to smooth inhomogeneities. In such case, the connection between protected ground state degeneracy and nonlocality becomes less obvious. One can have zero modes with overlapping wavefunctions. The question remains as to whether the MZMs in smoothly inhomogeneous nanowires are susceptible to {parity-preserving} local perturbations. The specific type of perturbation and the spinorial internal structure $\mathbf{u}^{L,R}(x)$ become crucial in this regard. They lead to different forms of the susceptibility, expressed as overlap spatial integrals that give a sense of nonlocality, but in which internal degrees of freedom combine in different ways. In this subsection we analyse these different forms.

To make connection with published literature, we start by considering the response to global perturbations to the chemical potential $\mu$ in the nanowire. In Ref.~\cite{Ben-Shach:PRB15} it was shown that the zero-temperature change in the energy $E_0$ of a subgap Andreev state $c$ in response to such change 
is the state's dimensionless charge $\partial E_0/\partial\mu = \delta N =|Q|/e=\langle c^\dagger c-c c^\dagger\rangle$, which in the Majorana basis takes the form of a wave-function overlap
\begin{equation}
\delta N = \left|\int dx\,\mathbf{u}^L(x)\cdot\mathbf{u}^R(x)\right|.
\label{deltaN}
\end{equation}
Here $|\dots|$ denotes the absolute value. It becomes clear that for well separated, exponentially decaying Majorana wavefunction $\mathbf{u}^{L,R}(x)$ at a distance $L$, as corresponds to the MZMs of uniform topological nanowires, this susceptibility $\delta N$ is exponentially suppressed, $\delta N\sim e^{-L/\xi}$, just like $E_0$.

{We next consider the susceptibility of the Majoranas to a local, spatially uncorrelated noise source that preserves fermion parity
\begin{eqnarray}
W_{P}(t) &=& \int dx \, W_{P}(x,t) \nonumber\\
&=& \int dx\,\sum_{\sigma\sigma'}\psi^\dagger_\sigma(x)P_{\sigma\sigma'}\psi_{\sigma'}(x)V(x,t),
\end{eqnarray}
where $P$ is a certain operator on spin space and $V(x,t)$ a spatially uncorrelated random energy fluctuation, with a  power spectral density $\overline{V(x,t)V(x',t')} = \delta(x-x')S(t-t')$. Examples of such perturbations include short-range fluctuations of the electrostatic potential (with $P=P_0=\delta_{\sigma\sigma'}$), a Zeeman-like perturbation due to exchange with an ensemble of local magnetic impurities $\vec{m}(x)$ polarised along the Zeeman field (with $P=P_3=\sigma_z$, the $z$ Pauli matrix), or completely spin-uncorrelated local noise (with $P=P_\pm=[P_0\pm P_3]/2$), acting independently on opposite spins $\sigma=\pm 1$. More general models not considered here, such as multimode nanowires, can be considered within this same formalism by including any additional quantum numbers (e.g. transverse modes) into $\sigma$, and defining the corresponding $P$ matrices for relevant (possibly mode-mixing) perturbations. We note that without mode-mixing noise, the results obtained in this work would apply directly to multimode nanowires too.}

{A given perturbation $W_{P}$ produces dephasing within the zero-energy Majorana subspace, with the dephasing rate proportional to the susceptibility of $E_0$ to $W_{P}$ at vanishing frequency, $1/T_2^*\propto\int dx\left|\partial E_0/\partial W_{P}(x, \omega=0)\right|$. This relation follows from e.g. the analysis in Ref.~\onlinecite{Knapp:PRB18}, generalised to include statistically independent local noise sources (note the absolute value inside the spatial integral, as expected for the addition of independent dephasing channels). Using first-order perturbation theory, $\partial E_0/\partial W_{P}(x,\omega=0) \propto \sum_{\sigma\sigma'}u_\sigma^L(x)P_{\sigma\sigma'}u_{\sigma'}^R(x) \equiv \chi_P(x)$. The susceptibility $\chi_P(x)$ thus connects dephasing to various forms of spatial overlaps, $1/T_2^*\propto \Omega_k$, where $\Omega_k = \int dx\,|\chi_{P_k}(x)|$ is in fact an integrated susceptibility.}

{For completely arbitrary, local, spin-uncorrelated noise $W_{P_\pm}$
one obtains an integrated susceptibility $\Omega_s$ of the form
\begin{equation}
\Omega_s = \Omega_{+} + \Omega_{-} = \int dx \sum_\sigma\left|u_\sigma^L(x)u_\sigma^R(x)\right|.
\label{OmegaS}
\end{equation}
Despite the superficial similarity to $\delta N$ in Eq. \eqref{deltaN},  the spin degree of freedom enters differently in this case}. We note that this expression is not SU(2) symmetric, as assuming uncorrelated spin fluctuations requires one to define a preferred spin quantization axis. It is the relevant susceptibility when {the SU(2) symmetry of the system is broken by e.g. a finite spin-orbit coupling and/or an external Zeeman field, as happens in a Majorana nanowire}, and one does not know anything about the noise produced by the environment. It is therefore the `pessimistic' form of the susceptibility, which again takes the form of a measure of nonlocality. This measure is stricter than $\delta N$, since $\Omega_s\geq \delta N$.

If we consider only spin-independent local noise {($P=P_0$), i.e. short-range electrostatic potential fluctuations,} 
we have the following $\Omega_0$ susceptibility instead,
\begin{equation}
\Omega_0 = \int dx \left|\mathbf{u}^L(x)\cdot\mathbf{u}^R(x)\right|.
\end{equation}
Since this noise is more restricted, it is natural that $\Omega_0\leq\Omega_s$. It is thus a less strict measure of nonlocality than $\Omega_s$. It is moreover SU(2) symmetric.

Finally, we can define a degree of nonlocality that is purely spatial, and independent of the spin degree of freedom, defined as $1-\Omega_\mathrm{max}$, with
\begin{equation}
\Omega_\mathrm{max} \equiv \int dx \|\mathbf{u}^L(x)\|\, \|\mathbf{u}^R(x)\|.
\label{Omegamax}
\end{equation}
This Majorana overlap does not seem to be directly related with a linear-response susceptibility to any type of noise, but has the interesting property of being a strict upper bound to all other definitions. Taking into account Eq. (\ref{normalisation}), we have
\[
0\leq\delta N\leq\Omega_0\leq\Omega_s\leq\Omega_\mathrm{max}\leq 1.
\]
Remarkably, we will show that all these measures of nonlocality become essentially equal for zero modes in globally topological nanowires $B>B_c$, but not so for zero modes below the $B_c$ in inhomogeneous nanowires. This reflects the intricate relation between spin, nonlocality and protection derived from spin-orbit-induced spin textures.

Throughout this work we will deal in particular with the problem of estimating the most pessimistic susceptibility $\Omega_s$ with a local probe. This was the subject of theoretical and experimental studies in quantum dot-nanowire setups, see Refs.~\cite{Prada:PRB17, Deng:PRB18}. In this context, $\Omega_s$ is the pertinent measure of nonlocality, and given its meaning as a susceptibility to unrestricted local perturbations, it is also the most conservative measure of Majorana zero mode protection, beyond topological considerations. 

A trivial example of a MZM is a highly local Yu-Shiba-Rusinov state \cite{Yu:APS65,Shiba:POTP68,Rusinov:JL69,Lee:NN14} tuned to zero energy. In general, such a fully local Andreev state will have equal Majorana components $|u_\sigma^L(x)| = |u_\sigma^R(x)|$, and hence $\Omega_s=1$.  
A MZM of topological origin will in contrast have exponentially small overlap $\Omega_s\approx 0$, as e.g. the conventional $B>B_c$ topological zero modes in very long Majorana nanowires. As will become apparent in the course of this work, a continuum of ABSs are possible under broken time reversal symmetry with any value of $\Omega_s$ between zero and one. The so-called pseudo-MZMs at smooth interfaces will be shown to lie at any point {within this range $0\leq\Omega_s\leq 1$} depending on nanowire parameters.

\begin{figure*}
   \centering
   \includegraphics[width=0.9\textwidth]{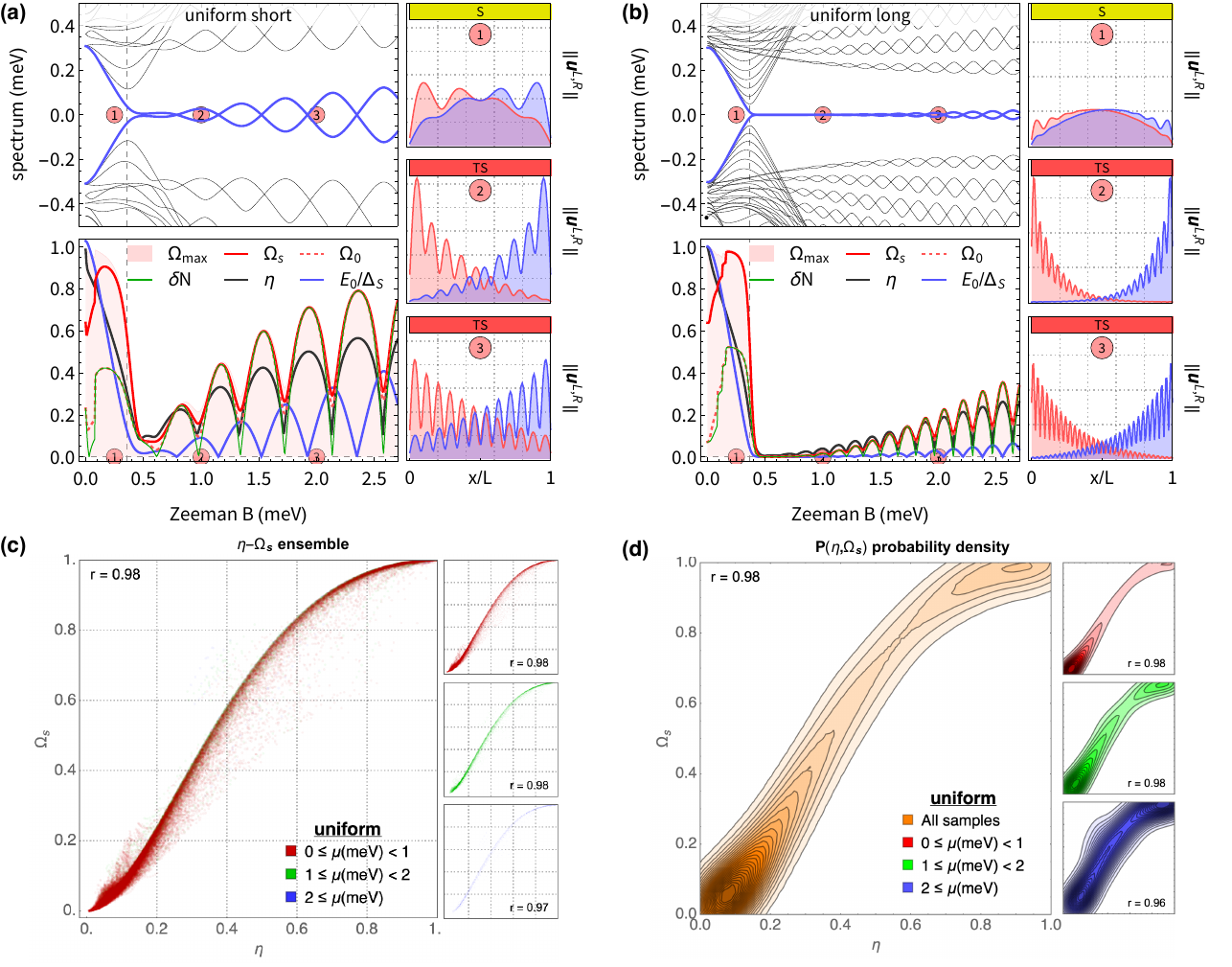}
   \caption{\textbf{Uniform nanowires.} Panels (a) and (b) show the spectrum (top panel), and the Majorana overlaps $\Omega_\mathrm{max}$ (red shaded region), $\Omega_s$ (solid red), $\Omega_0$ (dotted red), $\delta N$ (solid green), local estimator $\eta$ (black) and splitting $E_0$ (blue) of the lowest lying state (bottom panel) as a function of Zeeman $B$ in uniform nanowires with $\mu=0.2$ meV, $\Delta=0.3$meV and $\alpha=0.4$eV$\mathrm{\AA}$. The nanowire length is $L=1.2\mu$m (a) and $L=3\mu$m (b). The Majorana wavefunctions along the nanowire are shown to the right, at three values of $B$ (numbered circles), together with a sketch (above) of the corresponding band-topological phase of the nanowire. In panel (c) we show the position in the $(\eta,\Omega_s)$ plane of the near zero modes with $E_0\leq 10\mu$eV in an ensemble of nanowire configurations  uniformly distributed within the following parameter ranges: $\mu\in(0,2.5)$ meV, $\Delta\in[0,0.5]$ meV, $\alpha\in[0.1,1]$ eV$\mathrm{\AA}$ and $L\in[0,2] \mu$m. Panel (d) shows the corresponding probability density $P(\eta,\Omega_s)$. The Pearson correlation factor of the nonlocality estimator is $r=0.98$ [slightly less as we filter into bins of increasing Fermi energy, see small red, green, blue subpanels to the right of (c) and (d)].}
   \label{fig:uniform}
\end{figure*}

\subsection{Local estimator $\eta$}

It was proposed~\cite{Prada:PRB17} to estimate $\Omega_s$ by relating it to a quantity $\eta$ that can be extracted by a local measurement performed at $x=0$. This point is chosen here as the left end of the nanowire, of total length $L$. Other choices for $x$ can be considered, but in our models at least one of the zero modes will often be concentrated around said end, so this becomes the optimal choice for our purposes, and the one relevant for most experiments currently. As discussed below, the measurement itself can be of different kinds, but it should be designed to probe a local quantity $\eta$, intrinsic to the isolated nanowire, defined in terms of the ratio of the norms of the two Majorana components at $x=0$,
\begin{equation}
\eta \equiv \sqrt{\frac{\|\mathbf{u}^R(x=0)\|}{\|\mathbf{u}^L(x=0)\|}}.
\label{eta}
\end{equation}

The original observation that this quantity is correlated with Majorana nonlocality was made in Ref. \cite{Prada:PRB17}, where it was shown that in $B>B_c$ uniform nanowires
\begin{equation}
\Omega_s\approx \eta.
\end{equation}
In the rest of this work we will quantitatively evaluate how well the approximation holds in more general situations, including inhomogeneous samples with smooth interfaces hosting near-zero modes at $B<B_c$.

At least two schemes have recently been proposed to access the local quantity $\eta$. The first~\cite{Avila:A18} consists in measuring transport from a normal tunneling probe, coupled to the nanowire through a barrier at $x=0$. By analysing the profile of the zero-bias anomaly associated to the two Majoranas, one may extract their respective decay rates $\Gamma_L$ and $\Gamma_R$~\cite{Ioselevich:NJP13}. These rates can be expressed as effective coupling amplitudes $t_L, t_R$ of the two Majoranas across the barrier, $\Gamma_{L,R}=\pi \rho_0 |t_{L,R}|^2$, which can in turn be expressed in terms of overlap integrals of probe and Majorana wavefunctions under the barrier using Bardeen's tunneling theory~\cite{Bardeen:PRL61}. For a short and high barrier, the ratio of such overlap integrals are given by the intrinsic $\|\mathbf{u}^{R}(x=0)\|/\|\mathbf{u}^{L}(x=0)\|$ in the decoupled nanowire, so that $\eta$ can be approximated by 
\begin{equation}
\eta \approx \left({\frac{\Gamma_R}{\Gamma_L}}\right)^{1/4}.
\end{equation}
The second scheme~\cite{Prada:PRB17,Clarke:PRB17}, recently demonstrated experimentally~\cite{Deng:PRB18}, consists in measuring, using tunnelling spectroscopy, the splitting of the zero mode as it is tuned to resonance with a quantum dot state coupled in series to the end of the nanowire (Fig. \ref{fig:sketch}). This scheme, applied to two subsequent dot-wire resonances, gives access to the couplings $t_L$ and $t_R$ of the two Majoranas to the dot, in terms of which $\eta$ reads
\begin{equation}
\eta \approx \sqrt{\frac{|t_R|}{|t_L|}}.
\end{equation}
Again, this approximation to the intrinsic $\eta$ is valid for short barriers between dot and nanowire. The dot measurement scheme also yields information about the spin-canting  angles of the zero modes at $x=0$~\cite{Prada:PRB17,Schuray:A18}. We will return to the spin of the MZMs in Sec. \ref{sec:spin}, after analysing in detail $\eta$ as an estimator of $\Omega_s$.

We first note that a MZM of topological origin, with $\Omega_s\approx 0$, is guaranteed to have $\eta=0$, as $\Omega_s \approx 0 \Rightarrow u_\sigma^R(0)\approx 0$ (though the converse is not true). Likewise, a perfectly local ABS with $\Omega_s=1$ requires $u_\sigma^L(x) = u_\sigma^R(x)$ for all $x$, so $\eta=1$ in that case. The correlation between $\Omega_s$ and $\eta$ in intermediate situations is not so simple, and depends on the specific microscopic configuration of the Majorana nanowire.

We will explore three such types of configurations: (1) a uniform Lutchyn-Oreg nanowire, (2) a superconducting nanowire with a smooth step in the electronic density (S'S), and (3) a nanowire with a step both in charge density and induced pairing (NS). Within this latter class we further specialise in two specific cases of particular experimental relevance, a superconducting nanowire with a narrow normal barrier (4) or a quantum dot (5) at the left end of the nanowire (see sketch in Fig. \ref{fig:sketch}). These five setups, each corresponding to a different  device design, play an important role in the ongoing discussion around the interpretation of recent experimental observations and even of theoretical results themselves. They will now be discussed in turn.

Before proceeding, it is interesting to make the connection between the local estimator $\eta$ and the non-Hermitian topological classification theory mentioned in the introduction. Zero modes with intermediate $\Omega_s$ acquire a distinct topological classification when coupled to a reservoir at $x=0$~\cite{Avila:A18}. Within this theory, any deviation from perfect locality $\eta<1$ translates into a dimensionless asymmetry in the couplings of each Majorana component to the reservoir, denoted as $\gamma_0/\Gamma_0$, where $\gamma_0$ and $\Gamma_0$ are, respectively, the half-difference and average of the escape rates of the two Majoranas into the reservoir. The connection with $\eta$ is simply $\eta^4=(1-\gamma_0/\Gamma_0)/(1+\gamma_0/\Gamma_0)$. A finite asymmetry, in turn, stabilises the zero mode through an exceptional point bifurcation. This happens if the coupling asymmetry exceeds the Majorana splitting $\gamma_0>|E_0|$. Hence, a  mode with $\eta<1$ \emph{exactly} at zero energy will be rendered non-trivial when coupled to a reservoir.

\section{Uniform Majorana nanowires}
\label{sec:uniform}

The established standard to describe Majorana nanowires is the Lutchyn-Oreg model \cite{Lutchyn:PRL10, Oreg:PRL10}. It consists in a modification of an original proposal by Fu and Kane \cite{Fu:PRL08} to engineer one-dimensional topological superconductivity by proximity to a conventional superconductor. The ingredients of the Lutchyn-Oreg model are a one dimensional semiconducting nanowire with Fermi energy $\mu$ and spin-orbit coupling $\alpha$, a Zeeman field $B$ applied parallel to it, and an s-wave superconductor to induce a pairing $\Delta$ on the nanowire by proximity effect. The  Bogoliubov-de Gennes Hamiltonian of the model reads
\begin{equation}
H = \left(\frac{p_x^2}{2m^*} - \mu\right) \tau_z + B\sigma_x\tau_z - \frac{\alpha}{\hbar} p_x \sigma_y \tau_z + \Delta\tau_x.
\label{H}
\end{equation}
Here $\tau_i$ and $\sigma_i$ are Pauli matrices in the $(c^\dagger, c)$ particle-hole and $(\uparrow, \downarrow)$ spin sectors. Only one spinful nanowire subband is included. We use, for concreteness, the effective mass of InSb, $m^*=0.015m_e$. The model describes a trivial superconducting phase for $B<B_c\equiv \sqrt{\mu^2+\Delta^2}$, while for $B>B_c$ it develops a topological p-wave gap with Majorana zero modes at either end of a long nanowire. For smaller lengths $L$ around one micron, Majoranas start to develop a finite overlap $\Omega_s$ and hibridise away from zero energy $E_0>0$.

This model has been extensively used to characterise the basic physical regimes of Majorana nanowires. The correlation between $\Omega_s$ and the estimator $\eta$ established in Ref. \onlinecite{Prada:PRB17} referred mainly to this model. Here we use it as a starting point for the more complicated inhomogeneous nanowire models discussed later. We computed its associated phenomenology using the MathQ package \cite{MathQ}. Figure \ref{fig:uniform} summarises the results. Panel (a) shows, for a $L=1.2$ $\mu$m nanowire, the $B$ dependence of the Bogoliubov spectrum, the typical wavefunctions, their hybridisation energy $E_0$, their overlaps $\delta N$, $\Omega_0$, $\Omega_s$ and $\Omega_\mathrm{max}$, and the overlap estimator $\eta$. Panel (b) shows the equivalent results for a longer $L=3$ $\mu$m nanowire.

The spectra in panels (a,b) illustrate the well-known band inversion at a finite $B_c\approx 0.36$ meV (dotted vertical line), after which a zero mode emerges that oscillates as a function of $B$ due to the small, but finite, spatial overlap of its Majorana components. The corresponding Majorana wavefunctions $\|\mathbf{u}^{L,R}(x)\|$ for the lowest eigenstate are depicted for three values of $B$ (numbered circles), one below $B_c$ and two above. Atop the wavefunctions we represent the corresponding band-topological phase along the nanowire, with trivial in yellow (S) and topological in red (TS). Note that for field $B_1=0.25\mathrm{meV}<B_c$, although the topological transition has not yet taken place, the Majorana components of the lowest eigenstate appear as precursors to the Majorana zero modes at the higher $B_2=1.0 \mathrm{meV}>B_c$ and $B_3=2.0 \mathrm{meV}>B_c$. 
Their overlap ceiling $\Omega_\mathrm{max}$ (red shaded region) starts from $\Omega_\mathrm{max}=1$ (strictly local) at $B=0$, and decreases as $B$ increases.  
We see that the wave-function overlap between MZMs is a continuous non-monotonic quantity as a function of Zeeman field (this will also be the case in inhomogeneous nanowires). Thus, it is incorrect to think about fully local/nonlocal MZMs, before/after a topological transition. Interestingly, in uniform nanowires the spin-independent forms of the overlaps coincide for $B<B_c$, $\delta N\approx\Omega_0$, while the overlap corresponding to spin-uncorrelated noise essentially follows the fully spatial overlap, $\Omega_s\sim\Omega_\mathrm{max}$. All forms of the overlap reach a common minimum just beyond the critical field $B\gtrsim B_c$, and then \emph{grow} together in an oscillatory fashion~\cite{Prada:PRB12,Das-Sarma:PRB12, Mishmash:PRB16, Dominguez:NQM17}, out of phase with the splitting (blue curve). Namely, MZMs deep in the topological regime exhibit \emph{less nonlocality} than the ones near $B_c$ (compare wavefunctions at $B_2$ and $B_3$). 

\begin{figure*}
   \centering
   \includegraphics[width=0.9\textwidth]{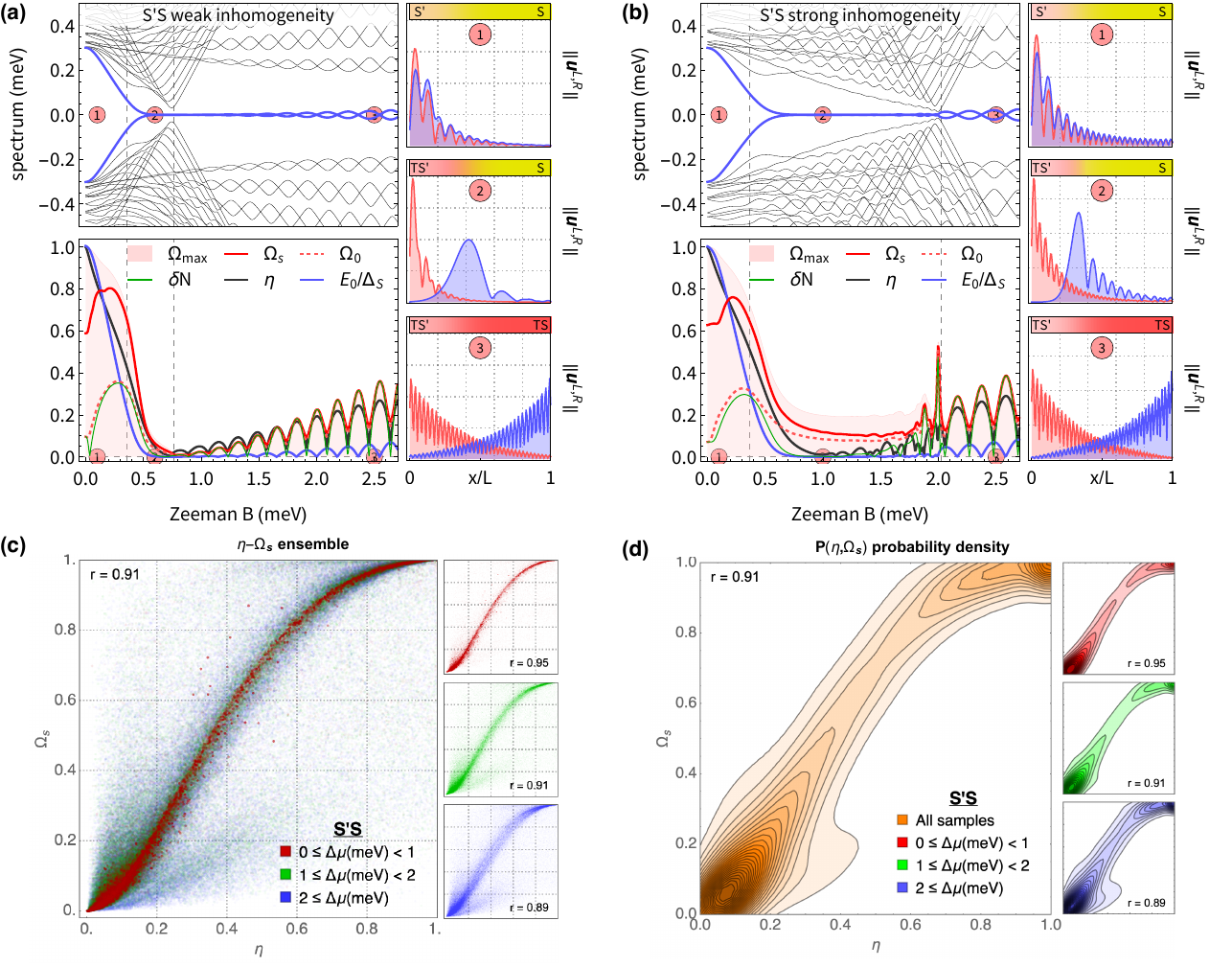}
   \caption{\textbf{Smooth S'S nanowires}. (a,b) Equivalent of Fig. \ref{fig:uniform} in a nanowire with two superonducting halves, of lengths $L_{S'}=1\mu$m and $L_{S}=2\mu$m, with uniform pairing $\Delta=0.5$ meV but different $\phi_{S'}=-0.2$meV and $\phi_S=-0.7$meV (a) or $\phi_S=-2.0$meV (b) at either side. In terms of Fermi energy differences $\Delta\mu=\max_x[\phi(x)]-\min_x[\phi(x)]$, this corresponds to $\Delta\mu=0.5$ meV (a) and $\Delta\mu=1.8$ meV (b).  The interface has smoothness length $\zeta=0.5\mu$m. The rest of parameters as in Fig. \ref{fig:uniform}.  Panels (c,d) are computed by independently sampling all parameters, $B\in[0,2.7]$ meV, $\Delta\in[0,0.5]$ meV, $\phi_{S',S}\in[-2.5,2.5]$ meV, $L_{S',S}\in[0.2,1]\,\mu$m, $\zeta\in[0,0.5]\,\mu$m and $\alpha\in[0.1,0.7]$ eV$\mathrm{\AA}$. The binning windows are defined in terms of $\Delta\mu$.}
   \label{fig:S'S}
\end{figure*}

In the two uniform nanowires simulated in panels (a,b) we see how the estimator $\eta$ (black curve) roughly traces the overlap $\Omega_s$, particularly around the $B\gtrsim B_c$ region of minimal overlap. 
This good correlation corresponds to two particular configurations of the uniform Lutchyn-Oreg model. To fully assess the overall correlation between $\eta$ and $\Omega_s$ for arbitrary configurations, we simulate an ensemble of $\sim 5\cdot 10^5$ uniform nanowires with varying model parameters, including $B$, $\alpha$, $L$, $\mu$ and $\Delta$, distributed uniformly within realistic ranges (see caption). Amongst all configurations, we select those with a near zero mode (splitting $E_0<10\mu$eV) well separated from higher excitations (second eigenvalue greater than $50\mu$eV). This preselection is experimentally feasible at temperatures below $\sim 100$mK using local tunnel spectroscopy. In the uniform case it excludes in particular any field in the trivial regime $B<B_c$. We then compute the $(\eta, \Omega_s)$ pair for the near-zero modes in the ensemble and collect all these points (see Fig. \ref{fig:uniform}c) to build the probability density $P(\eta,\Omega_s)$ that a given zero mode with a measured $\eta$ have a given $\Omega_s$. The probability $P(\eta,\Omega_s)$ is shown in Fig. \ref{fig:uniform}d. Its profile gives an accurate account of the quality of $\eta$ as an estimator of $\Omega_s$ within the whole space of uniform Lutchyn-Oreg nanowire models with near zero modes. A perfect correlation would appear as a straight, thin $P(\eta,\Omega_s)$ along the diagonal. We see that while the actual dependence of the typical $\Omega_s$ with $\eta$ is non-linear, the probability distribution is rather narrow and close to the diagonal, which reveals the high quality of the estimator within this model space. The Pearson correlation coefficient between $\eta$ and $\Omega_s$ is $r=0.98$. The small panels in red, green and blue dissect the ensemble according to their $\mu$ (see legend, low densities in red, higher densities in blue). The depicted partial $P(\eta,\Omega_s)$ show that the precision of $\eta$ is greater if the nanowire is known to have low electronic density. This will be a recurring fact throughout this study. Performing the same statistical analysis with $\Omega_0$ instead of $\Omega_s$ yields very similar results.

\section{Smooth S'S nanowires}
\label{sec:S'S}

In the remaining sections we consider inhomogeneous nanowires, described  by a generalised Lutchyn-Oreg model with position dependent pairing $\Delta(x)$ and electrostatic potential $\phi(x)$,
\begin{equation}
H = \left(\frac{p_x^2}{2m^*} -\mu+\phi(x)\right) \tau_z + B\sigma_x\tau_z - \frac{\alpha}{\hbar} p_x \sigma_y \tau_z + \Delta(x)\tau_x.
\label{H}
\end{equation}
(In what follows we reabsorb $\mu$ into $\phi(x)$ for simplicity.) 

Much of the current debate as to the potential non-triviality of transport signatures in Majorana nanowires revolves around the possibility that near zero modes may arise as the result of smooth spatial variations in $\Delta(x)$ and/or $\phi(x)$ in inhomogeneous nanowires, independently of a band-topological phase transition. The debate has thus centered mostly on distinguishing between topological MZMs and such pseudo-MZMs in this system, {with the implicit assumption that these states are fundamentally different in some sense. In our view, as summarised in the introduction, 
the only meaningful distinction between zero modes has to be based on their respective wave-function structure, which underlies in particular the defining property of MZMs, namely their resilience against parity-preserving perturbations.}
 (see Sec. \ref{sec:nonlocality}). Regardless of their connection to band-topology, smoothly-confined pseudo-MZMs with a sufficiently small overlap $\Omega$ will therefore be, for all purposes, genuine MZMs protected against the corresponding type of local perturbation, exactly like the topological $B>B_c$ MZMs in finite-length nanowires. The debate is thus reduced to clarifying if smoothly confined near-zero modes can have significantly suppressed overlaps.

Two relevant types of smooth variations are possible within the Lutchyn-Oreg model, smooth S'S and smooth NS boundaries. In this section we concentrate on the S'S case, wherein $\Delta$ is uniform along the nanowire but $\phi(x)$ is position dependent. It can be positive (insulating regions) or negative (higher density regions). The spatial variation may arise due to e.g. non-uniform screening from contacts or gates. We model $\phi(x)$ in a nanowire spanning $0<x< L_{S'}+L_S$ as 
\begin{equation}
\phi(x) = \phi_{S'} + (\phi_{S}-\phi_{S'})\theta_\zeta(x-L_{S'})
\end{equation}
where $\theta_\zeta(x) = \frac{1}{2}[1+\tanh(x/\zeta)]$ is a smooth step function of width $\zeta$, see Fig. \ref{fig:sketch}. This length controls the smoothness of the boundary between the left $S'$ side, of length $L_{S'}$ and the right $S$ side, of length $L_S$.

Similarly to the uniform nanowire, a non-uniform S'S system with sufficiently long $L_{S,S'}$ may still be analysed from the conventional point of view of band topology of the two sides. The two $\phi_{S'}$ and $\phi_S$ now define two critical fields $B_c^{S', S} = \sqrt{\Delta^2+\phi_{S', S}^2}$. For a given $B$, we can have all possible combinations S'S, TS'S, S'TS, and TS'TS, where S stands for a trivial superconductor, and TS a topological superconductor, depending on whether $B<B_c^{S',S}$ or $B>B_c^{S',S}$. Whenever the topology of the left and right sides is different (\emph{locally} topological nanowire) and the corresponding halves are long enough, a pseudo-MZM will be localised somewhere in the smooth junction, regardless of $\zeta$. This state is actually a consequence of the bulk boundary correspondence. There is therefore nothing `pseudo' about it. 
Crucially, moreover, we will show in the next section that this state is essentially identical to the so-called pseudo-MZMs of smooth NS junctions, where band-topological arguments do not apply. We thus argue that it is incorrect to distinguish between MZMs and pseudo-MZM in general isolated systems, as the two types of states are ultimately connected. The discussion, once more, should focus instead on the overlap $\Omega_s$, not on artificial distinctions between classes of zero modes.

Figure \ref{fig:S'S}, analogous to Fig. \ref{fig:uniform}, shows the overlaps, $\eta$ and spectral phenomenology of a smooth S'S nanowire as depicted in case (2) of Fig. \ref{fig:sketch}.  When the junction is sufficiently smooth, the S'S to TS'S transition at $B=B_c^{S'}$ manifests as a single subgap state dropping into the gap towards zero energy. A lone subgap level detaching from the quasicontinuum of levels is a recurrent and distinct feature of smooth configurations that replaces the band inversion characteristic of uniform wires. It is clearly visible in the spectrum of panels (a) and (b), blue curve, where parameters are chosen so that $B_c^{S'} < B_c^S$ (the two critical fields are shown as dotted vertical lines). The two panels (a,b) correspond, respectively, to S'S nanowires with weaker and stronger inhomogeneity $\Delta\mu=0.5$ meV and $\Delta\mu=1.8$ meV, where $\Delta\mu\equiv\max_x(\phi(x))-\min_x(\phi(x))$ is the maximum variation of the Fermi energy in the nanowire. We again show the Majorana component wavefunctions of the lowest eigenstate at three fixed fields (numbered circles). At $B_1=0.1\mathrm{meV}<B_c^{S'}$ the nanowire is in an S'S configuration (trivial-trivial), and the finite energy state dropping into the gap is merely a precursor of the Majorana zero modes at larger fields, concentrated on the less dense $S'$ side. It already exhibits a slightly suppressed overlap $\Omega_s<1$, with its two Majorana components starting to separate [wavefunction (1)]. As the nanowire enters the TS'S configuration [$B_c^{S'}<B_2 = 0.6 \mathrm{meV}<B_c^S$, wavefunction (2)] the MZMs at the smooth junction (blue) moves away from the left Majorana at $x=0$ (red). The distance between the two is $B$-dependent, since the TS length of the nanowire that satisfies $B>\sqrt{\Delta^2+\phi(x)^2}$ [see colored bar atop wavefunction (2)] grows with $B$ due to the smooth $\phi(x)$ profile. The spatial decoupling suppresses $\Omega_s$ (solid red curve in bottom panel)] until $B_c^S$ is reached, wherein the type of $\Omega_s$ oscillations we observed in the uniform case appear, and the Majorana wavefunctions become conventional, confined to the ends of the nanowire [wavefunction (3)]. 

\begin{figure*}
   \centering
   \includegraphics[width=0.9\textwidth]{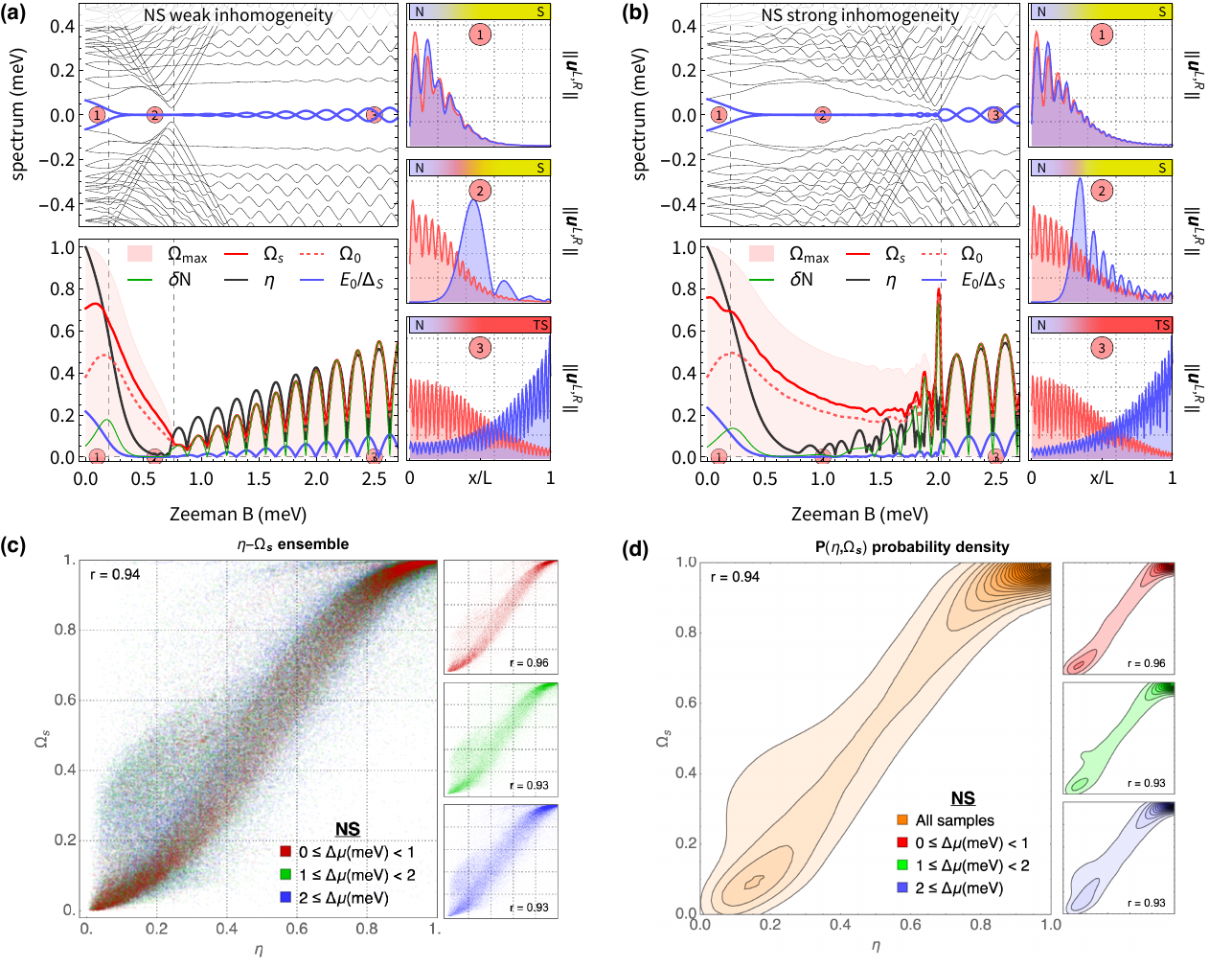}
   \caption{\textbf{Smooth NS nanowires}. Equivalent to Figs. \ref{fig:uniform} and \ref{fig:S'S}, with identical model and sampling parameters as in the latter, except for a zero pairing $\Delta_N=0$ on the left side and finite $\Delta_S=0.5$ meV on the right side of the smooth junction. Note the similar wavefunctions of the smooth junction Majoranas as compared to the S'S case of Fig. \ref{fig:S'S}.}
   \label{fig:NS}
\end{figure*}

Contrary to conventional lore, the Majorana overlap in the globally topological TS'TS phase at $B>B_c^{S,S'}$ is not necessarily smaller than in the locally topological TS'S case with a Majorana within the bulk of the nanowire. For example, 
$\delta N$ and $\Omega_0$ can be substantially suppressed for $B<B_c$, which suggests a strong resilience of locally topological Majorana zero modes against electrostatic potential fluctuations, see Figs. \ref{fig:S'S}(a,b), bottom panels.
Actually, as was noted also for the uniform case, in typical S'S nanowires shorter than around $L\sim 3\mu$m all forms of the Majorana overlap reach their minimum within the TS'S regime, $B_c^{S'}<B<B_c^{S}$, and begin to \emph{increase} into the TS'TS phase. The same will be noted in Sec. \ref{sec:NS} for smooth NS nanowires. This behavior is due to the different (faster) decay profile of $u_\sigma^R(x)$ when it lies at the smooth TS'S junction than when it shifts to the abrupt right boundary of the nanowire. It is important to appreciate the difference between these two types of MZMs. The MZM at a smooth boundary is also spatially smooth, with a Gaussian-like profile~\cite{Kells:PRB12,Fleckenstein:PRB18}, while the MZM at an abrupt boundary has fast $\sim k_F$ spatial harmonics and a double-exponential decay~\cite{Klinovaja:PRB12}. We will analyse in more detail the profiles and spin densities of these two types of MZMs in Sec. \ref{sec:spin}. In our current setup, this results in a smoother $B$ dependence of the overlap within the TS'S regime $B_c^{S'}<B<B_c^S$ as compared to the TS'TS regime $B>B_c^S$. 

The faster spatial decay of smooth Majoranas suggests that the accuracy of the local estimator $\eta$ should be worse in this case, compared to the case of uniform nanowires and abrupt MZMs. Indeed, the estimator may become suppressed as the smooth Majorana moves away from $x=0$ at a faster rate (Gaussian) than the overlap (exponential). We find that in realistic nanowires (see parameter ranges in the caption to Fig. \ref{fig:S'S}) the accuracy of $\eta$ is indeed reduced, particularly under strong inhomogeneities $\Delta\mu\gg\Delta$. This is shown in Fig. \ref{fig:S'S}(c,d). Here we have performed, once more, a sampling over nanowire parameters, this time including also $\phi_{S'}$, $\phi_S$, $L_{S'}$, $L_S$ and $\zeta$ [see Figs. \ref{fig:S'S}(c,d)]. The resulting $P(\eta,\Omega_s)$ is similar to that of the uniform case, albeit with a slightly reduced Pearson coefficient $r=0.91$. This effect is precisely the result of the Gaussian profile of smooth MZMs, which translates into a slight `bulge' above the origin and another one to its right. In the subpanels to the right we disect $P(\eta,\Omega_s)$ into partial probability densities for increasing degree of Fermi energy inhomogeneity $\Delta\mu$. We find that for inhomogeneities $\Delta\mu<1$ meV, the estimator preserves a high $r=0.95$ correlation with $\Omega_s$ (red subpanel), but increasing $\Delta\mu$ (green, blue subpanels) suppresses $r$, though the effect is not drastic, with $r\approx 0.9$ still. This remains true regardless of the maximum nanowire density considered.

\section{Smooth NS nanowires}
\label{sec:NS}

We now consider the second type of inhomogeneous nanowire, wherein the pairing, like $\phi(x)$, is also position dependent, $\Delta(x)$. We again consider a simple profile that interpolates between a left side and a right side. The left side is always normal in this case, with $\Delta_N=0$, so that the nanowire contains a smooth NS interface centered at $x=L_N$,
\begin{eqnarray}
\phi(x) &=& \phi_{N} + (\phi_{S}-\phi_{N})\theta_{\zeta}(x-L_{N}),\nonumber\\
\Delta(x) &=& \Delta_S \theta_{\zeta}(x-L_{N}).
\end{eqnarray}

This model is relevant to many devices explored in recent experiments. Nanowires are often made superconducting by growing an epitaxial superconductor on their surface. Often, the epitaxial coverage of the nanowire is incomplete, so it is natural to assume a suppressed pairing in the exposed portions. Like in the S'S nanowire, a thorough microscopic validation of this model would require a detailed characterisation of the device in question.

\begin{figure}
   \centering
   \includegraphics[width=0.95\columnwidth]{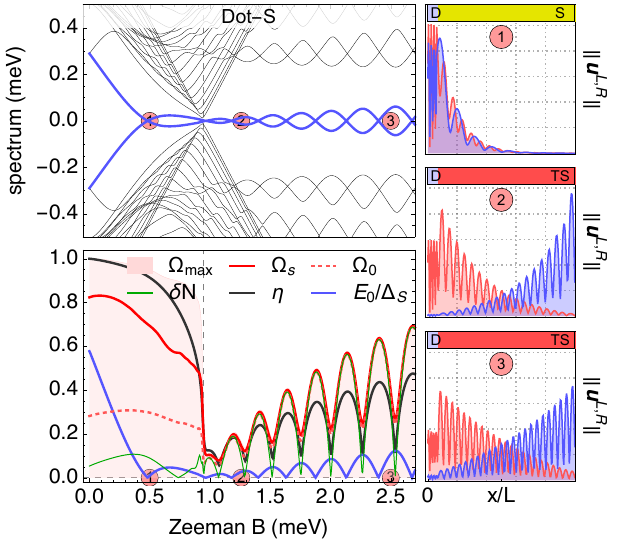}
   \caption{\textbf{Dot-Superconductor nanowires}. Specific Dot-S configuration of the NS model, relevant to a number of experiments~\cite{Deng:S16, Deng:PRB18}, with negative $\phi_{N}=-14$meV  and short $L_{N}=0.15\mu$m, which confines states in a quantum-dot region around $x=0$. Other parameters: $\phi_S=-0.8$ meV, $\Delta_S=0.5$ meV, $\zeta=0$, $L_S=2\mu$m and $\alpha=0.2$eV$\mathrm{\AA}$.}
   \label{fig:DS}
\end{figure}

The fundamental interest of the Lutchyn-Oreg model with a smooth NS interface is particularly high due to the fact that, perhaps surprisingly, it can also host near-zero modes at finite Zeeman field B, much like the smooth S'S, despite not developing a topological gap on the normal side. This is shown in Fig. \ref{fig:NS}, which is the NS version of Fig. \ref{fig:S'S}. The suppressed pairing gives rise to Andreev levels in the normal region. Depending on the normal length $L_N$, their level spacing 
$\delta\epsilon$ can be much smaller than the induced gap $\Delta$, which results is many subgap levels (unlike the S'S case, where only a lone level, detached from the quasicontinuum appears). A finite $B$ field Zeeman-splits all these subgap levels, that evolve avoiding each other due to spin-orbit coupling. This is true for all except the lowest two excitations (blue), which converge to zero energy with a \emph{finite} slope at low $B$-fields~\cite{Vaitiekenas:PRL18} (this is unlike in the S'S case, where the lone detached level starts off \emph{flat} at $B=0$)~\cite{Moor:NJP18}. Despite the superficial resemblance to Zeeman-induced parity crossings in quantum dots \cite{Lee:NN14,Moor:NJP18} (see Fig. \ref{fig:DS}), near-perfect Andreev reflection of N electrons on the smooth NS interface stabilises  this low-lying subgap level near zero energy for $B>\delta\epsilon$, but still well before $B_c^S$.  

From the point of view of its Majorana components, this near-zero mode is remarkably similar to the corresponding zero mode at the TS'S junction. Comparing the wavefunctions at field $B_2$ and $B_3$ in Figs. \ref{fig:S'S} and \ref{fig:NS}, we see that the essential difference between the S'S and NS cases lies in the abrupt Majorana component $\mathbf{u}^L
(x)$ on the left side (red wavefunction). In the NS case it is delocalised throughout the N region of length $L_N$, while in the TS'S case it is confined within a coherence length of the $x=0$ boundary. The smooth Majorana at the junction, however, is very similar and, remarkably, remains confined at the junction instead of decaying into the N side. We note that in the NS case this confinement is not the result of a bulk-boundary correspondence, as the left side is not gapped in the continuum limit, but of the high smoothness $\zeta$ of the boundary which enhances Andreev reflection. This observation hints at a deeper reason for the zero energy of smoothly confined states beyond topology, connected instead to the exact charge-conjugate symmetry of quasiparticles that undergo perfect Andreev reflection at an adiabatically smooth S boundary.

\begin{figure}
   \centering
   \includegraphics[width=0.95\columnwidth]{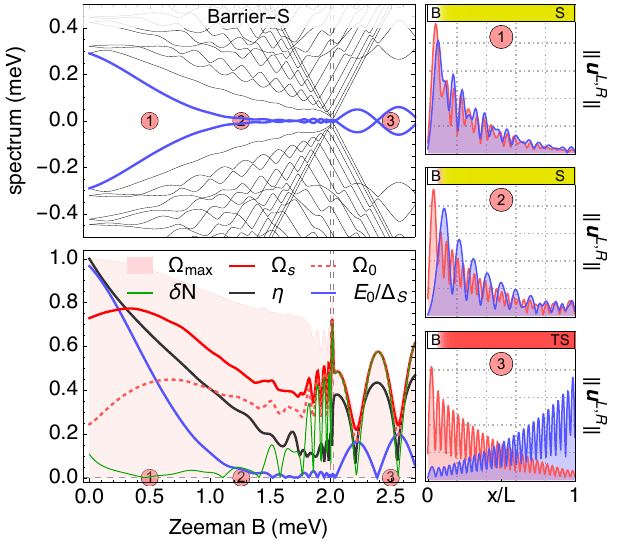}
   \caption{\textbf{Barrier-Superconductor nanowires}. Specific Barrier-S configuration of the NS model, similar to the one discussed in Ref. \cite{Vuik:18}, with positive $\phi_{N}=2$meV and short $L_{N}=\zeta=0.1\mu$m, which forms a smooth, Zeeman-polarised insulating barrier around $x=0$. Other parameters: $\phi_S=-1.8$ meV, $\Delta_S=0.3$ meV, $L_S=2\mu$m and $\alpha=0.4$eV$\mathrm{\AA}$.}
   \label{fig:BS}
\end{figure}

The similar wave-function phenomenology produces a similar behaviour also for $\Omega_s$ and $E_0$ as a function of $B$. Three regimes are visible, with crossovers at Zeeman $B\approx\delta\epsilon$ and $B\approx B^S_c=\sqrt{\Delta_S^2+\phi_S^2}$, see Fig. \ref{fig:NS}(a,b). In the regime $\delta\epsilon<B<B_c^S$ with smooth Majoranas, we see that $\eta$ underestimates the overlap $\Omega_s$.
The general $(\eta,\Omega_s)$ correlation analysis is shown in Fig. \ref{fig:NS}(c,d). The overall correlation, with identical sampling and zero-mode preselection scheme as in the S'S, is now $r=0.93$. This reduced value respect to the uniform case is once more due to the effect of the smooth MZMs. Due to their faster decay, their overlap with the delocalised $N$ Majorana components is greater than what $\eta$ would estimate. This produces the 'bulge' at $(\eta,\Omega_s)\approx (0.2,0.4)$ in the probability distribution, which is now more pronounced as compared to the S'S case. The states in this region  of underestimated overlap, however, come from highly inhomogeneous samples, as can be seen from the small subpanel decomposition. If the nanowire Fermi energy inhomogeneity $\Delta\mu$ is known to be low enough ($\Delta\mu<1\mathrm{meV}\approx 3\Delta_S$, red subpanel), the correlation remains strong at $r=0.96$.

\subsection{Smooth Barrier-S and Dot-S nanowires}
\label{sec:BDS}

A particular case of the NS nanowires in the preceding section that is of relevance to many devices is the limit in which $L_N$ is small. Nanowires designed to be probed by tunneling spectroscopy are often left uncovered by the superconductor at $x=0$ in order to allow efficient gating of the contact to the metallic reservoir. A finger gate under $x=0$ can then, thanks to the reduced screening by the superconductor shell, tune the transparency of the contact by inducing a positive $\phi_N$. This defines a barrier of finite smoothness $\zeta$, see case (4) in Fig. \ref{fig:sketch}. Such a setup was recently discussed in Ref. \onlinecite{Vuik:18}, where the smoothness allowed for the development of a stable $B<B_c$ near-zero mode, with a different coupling of its Majorana components across the Zeeman-polarised barrier  by virtue of their opposite spin orientation at the smooth contact. The phenomenology of a such a Barrier-S configuration is shown in Fig. \ref{fig:BS}. We see that the near-zero modes at the barrier for $B<B_c$ are characterised by a high overlap $\Omega_s$ but a reduced charge $e\delta N$ due to Andreev processes. 

An opposite voltage of the finger gate can make $\phi_N$ strongly negative. This may trap discrete states around $x=0$ in an effective quantum dot-superconductor configuration.  Additionally, screening effects in the nanowire may produce, in a mean-field approximation, a quantum dot-superconductor profile spontaneously \cite{Dominguez:NQM17, Escribano:BJN18} that can also trap states. To gain insight into these cases we simulate nanowires with short, normal dot regions abruptly connected to the nanowire ($\zeta=0$) without an additional intervening barrier, so the confinement is merely the result of the potential and pairing mismatch at $L_N$. This is a likely situation in experiments. Its associated phenomenology is shown in Fig. \ref{fig:DS}. The trapped states are Zeeman-split as $B$ is increased, and can cross zero energy at specific values of $B=B_1<B_c$ \cite{Lee:NN14,Moor:NJP18}, analogous to Shiba state parity crossings. The crossings are  considerably flattened due to the effect of Andreev reflections from the nanowire, which are enhanced by the lack of a confining dot-nanowire barrier. The near-zero mode is not completely stabilised at zero, unlike in Fig. \ref{fig:NS}, because Andreev reflection is however not perfect (that requires a smooth dot-S contact). The state remains very concentrated within the quantum dot region, and is therefore considerably local, with $\Omega_s$ and $\eta$ both close to one. Its charge $e\delta N$ and susceptibility $\Omega_0$ to local potential fluctuations are comparatively suppressed, again due to Andreev processes. This once more showcases the fact that seemingly trivial, spatially overlapping near-zero modes are not necessarily fragile, and may exhibit, due to Andreev particle-hole mixing, a highly non-trivial response to certain perturbations.\footnote{We note that introducing a barrier between dot and nanowire progressively suppresses these non-trivial Andreev effects until one reaches, for high barriers, a standard quantum dot behaviour with unitary charge and a conventional unprotected response to electrostatic and Zeeman noise.}

It is important to note that these Barrier-S and Dot-S types of configurations of the generic NS model are included in the NS sampling of Figs. \ref{fig:NS}(c,d), which therefore remains representative of the quality of the $\eta$ estimator expected in these cases. We have performed samplings of purely Dot-S and Barrier-S configuration ensembles, and found similar $P(\eta,\Omega_s)$ distributions as for the NS case, including the $(\eta,\Omega_s)\approx (0.2,0.4)$ bulge. The general conclusions on the NS model class can thus be applied also to Barrier-S and Dot-S models.

\section{Spin texture and smoothness}
\label{sec:spin}

\begin{figure}
   \centering
   \includegraphics[width=\columnwidth]{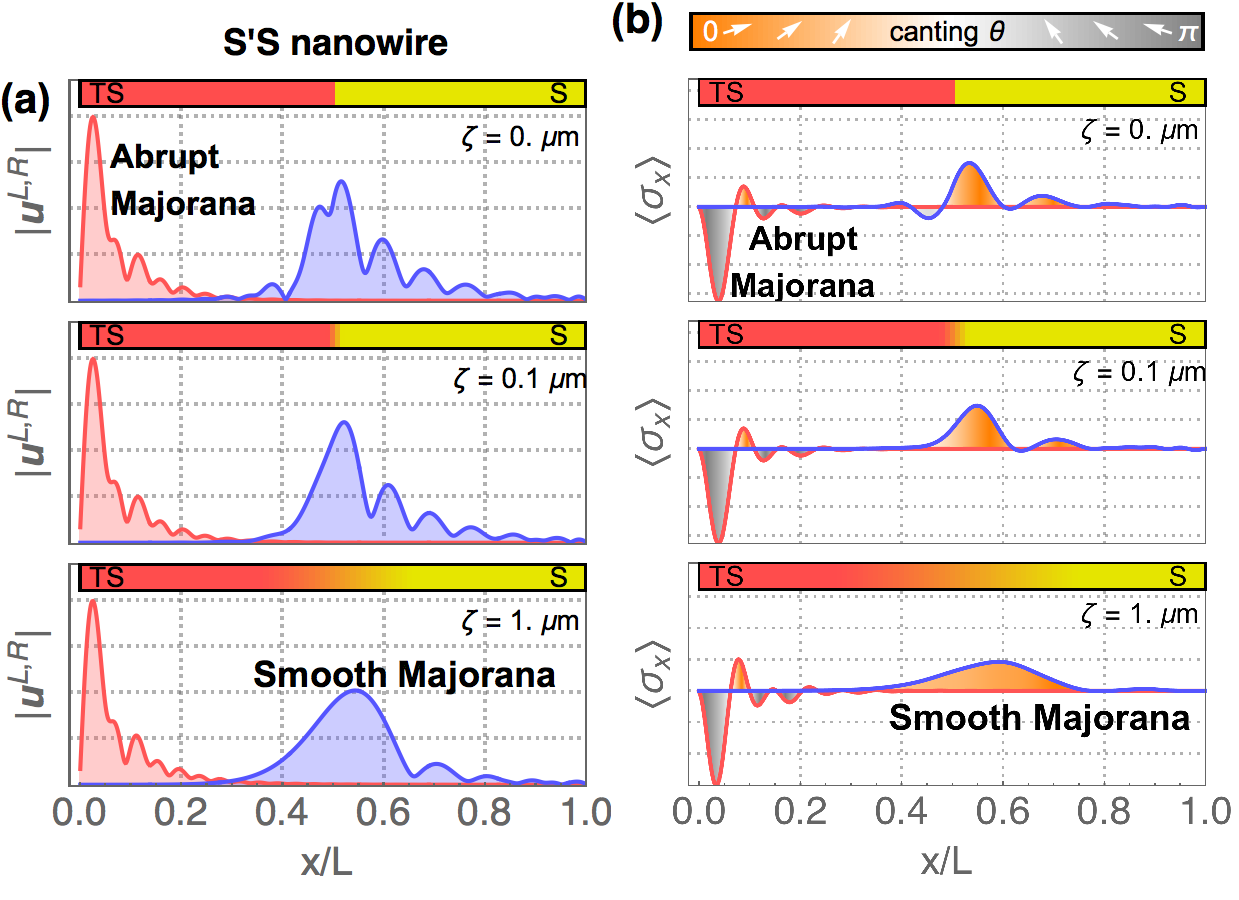}
   \caption{\textbf{Majorana spin in S'S junctions.} (a) Wavefunction $|\mathbf{u}^L|$ and $|\mathbf{u}^R|$ of the lowest eigenstate in an TS'S junction of increasing smoothness $\zeta=(0.0,0.1,1.0)\mu$m. (b) The corresponding spin density $\langle\sigma_x\rangle$ along the Zeeman field. The shading under the spin density curves encodes the Majorana canting angle $\theta$ relative to the Zeeman field along $x$. Parameters: $\phi_{S'}=0$, $\phi_{S}=-1$ meV, $L_{S'}=L_S=1.8\mu$m, $\Delta_{S'}=\Delta_S=0.4$ meV, $\alpha=0.4$eV$\mathrm{\AA}$.}
   \label{fig:szss}
\end{figure}

\begin{figure}
   \centering
   \includegraphics[width=\columnwidth]{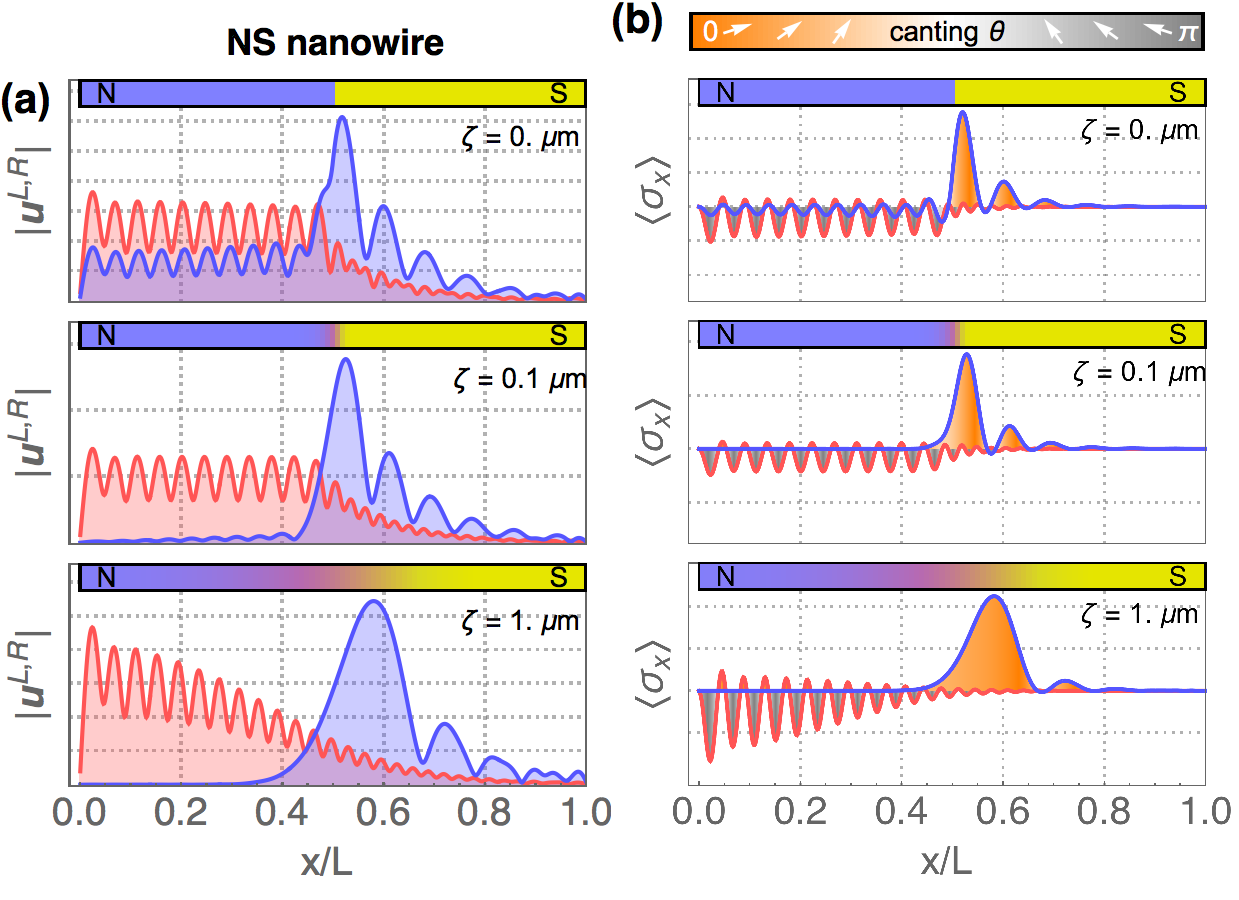}
   \caption{\textbf{Majorana spin in NS junctions.} Same as Fig.\ref{fig:szss} for an NS nanowire ($\Delta_N=0$).}
   \label{fig:szns}
\end{figure}

\begin{figure}
   \centering
   \includegraphics[width=\columnwidth]{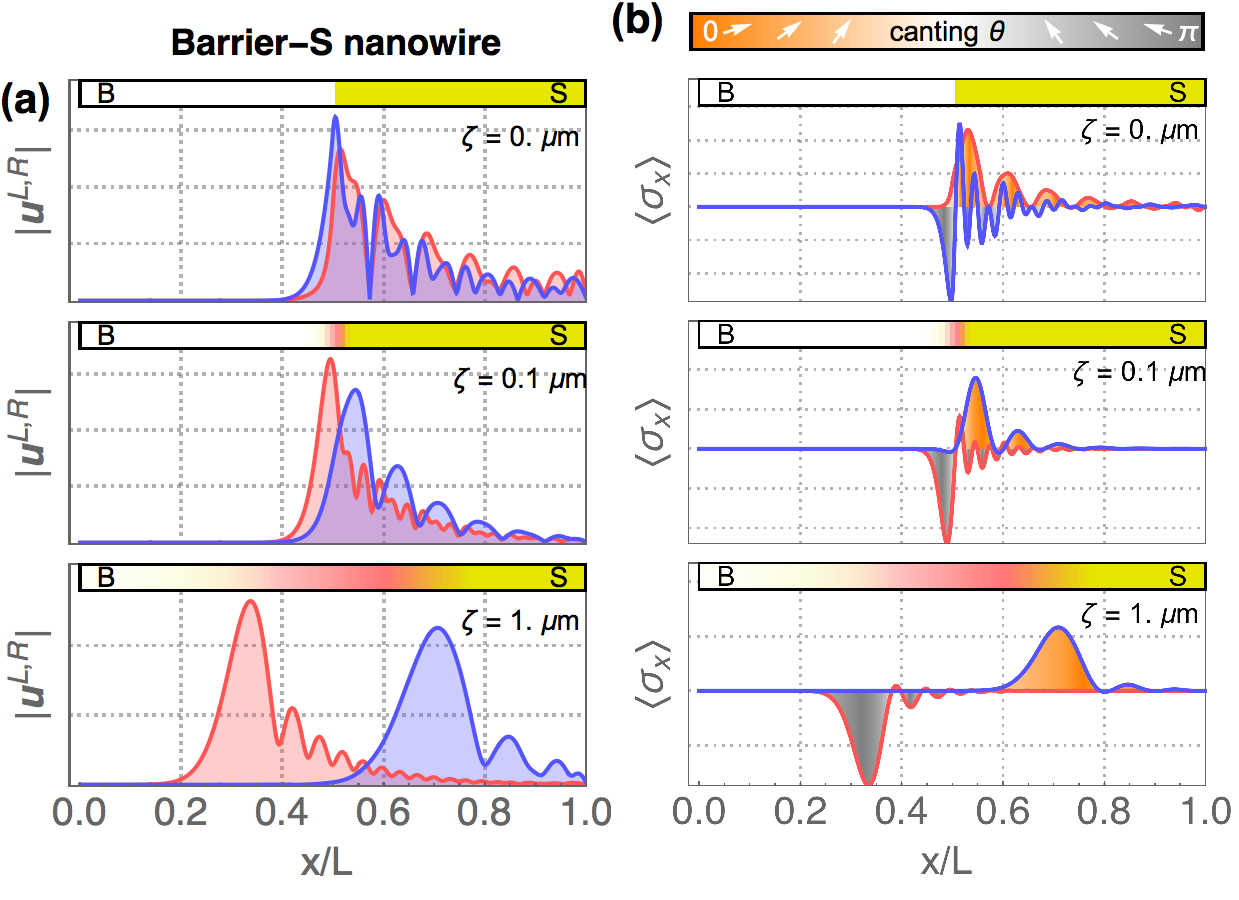}
   \caption{\textbf{Majorana spin in Barrier-S junctions.} Same as Fig.\ref{fig:szss} for an Barrier-S nanowire ($\Delta_N=0$, $\phi_N=1$ meV).}
   \label{fig:szbs}
\end{figure}

In this final section we analyse the spin structure of MZMs associated to different types of interfaces as a function of their smoothness. This aspect of the MZM wavefunction is relevant, since current experiments that extract $\eta$ to estimate the Majorana overlap use spin-polarised quantum dots coupled to the nanowire. The hybridisation of the dot levels and the MZMs at resonance depends strongly on the spin orientation of the latter. Furthermore, MZM spin is important in view of a recent arguments \cite{Vuik:18} that relate potential smoothness and spin polarisation. This work points out that at a smooth barrier, highly local MZMs acquire opposite spin polarisation, which may result in a highly  asymmetric coupling to a reservoir due to the Zeeman-polarisation of the barrier. We show here that such an effect is a particular manifestation of the MZM nonlocality produced by the barrier smoothness.

We once more analyse different nanowire configurations separately. Fig. \ref{fig:szss} shows, for an S'S nanowire of increasing smoothness $\zeta$ at $B_c^{S'}<B<B_c^S$, the Majorana wavefunctions of the lowest energy mode (a) and their spin polarisation $\langle \sigma_x\rangle$ along the nanowire (b). The first row shows a completely abrupt TS'S junction. The left Majorana $\mathbf{u}^L(x)$ centered at the abrupt $x=0$ boundary to vacuum (red curve) exhibits rapid oscillations associated to the Fermi wavevector $k_F$. We call this an \emph{abrupt} Majorana. Its fast spatial harmonics are the result of perfect $k_F\to -k_F$ normal reflection at the $x=0$ boundary. Its spin density likewise oscillates spatially within the $z-x$ plane. The angle $\theta$ in this plane (with $\theta=0$ for spin along $x$) is known as the canting angle, and is color coded in a gray-orange scale. The right Majorana, in blue, lies at the sharp TS'S interface where Andreev reflection processes are possible. It has a different profile from the abrupt Majorana, but still shows considerable density and spin oscillations. As the junction smoothness $\zeta$ increases (second and third row), the left Majorana remains unchanged, but the right Majorana at the  junction becomes increasingly smooth, loosing the fast spatial harmonics both in $\mathbf{u}^R(x)$ and $\langle\sigma_x\rangle$. Thus, a Majorana of Gaussian-like profile emerges, which we call here \emph{smooth} Majorana. Its spin becomes well defined, with canting angle converging to $\theta=0$ (orange) along the Zeeman field direction.

The equivalent smoothness phenomenology for the NS nanowire is shown in Fig. \ref{fig:szns}. In this case, the abrupt Majorana takes the form of a standing wave in the N region, with oscillatory density and spin. Its spin, however, is predominantly aligned along $-x$ (i.e. $\theta=\pi$, gray). The smooth Majorana, as remarked in Sec. \ref{sec:NS}, bears a strong resemblance to the one in smooth S'S junctions. It does not leak into the N side, even for moderate smoothness $\zeta\sim 0.1\mu$m, and acquires a well-defined spin polarisation along $x$ (i.e. $\theta=0$, orange). Again, the difference in density and spin texture of abrupt and smooth Majoranas in smooth nanowires is stark.

Finally, we present in Fig. \ref{fig:szbs} the results for a Barrier-S nanowire (insulating left side, $\phi_N>0$), with a barrier of increasing smoothness. For a sharp barrier (top row), the two Majoranas are very similar to the abrupt Majorana at $x=0$ in the S'S case. The only difference is that the barrier side has a finite potential, and a slight leakage of the two Majoranas is possible. The leakage, as pointed out in Ref. \cite{Vuik:18}, depends on the spin density of each Majorana, as the barrier height is different for the two spin orientations due to the uniform Zeeman field in the whole system, barrier included. Said spin orientation for the abrupt junction is rapidly varying, as corresponds to abrupt Majoranas. The difference in leakage becomes more pronounced as the barrier smoothness increases (middle and bottom rows). The spin of the two Majoranas in this case becomes increasingly well defined, and opposite, so that one Majorana penetrates more and more into the barrier as it becomes smoother. This leads to a simultaneous \emph{spatial and spin decoupling} (suppression of $\Omega_s$) of the two Majoranas at smooth barriers. We thus see that smoothness-induced nonlocality and spin-induced decoupling of Majoranas are one and the same. We conclude that, in the context of nanowires coupled to external reservoir~\cite{San-Jose:SR16,Avila:A18}, a different decay of MZMs into the outside world can always be traced back to a finite degree of nonlocality.

\section{Conclusion}
\label{sec:conclusion}

To summarise, in this work we have studied the properties of inhomogeneous Majorana nanowires. We have considered Majorana zero modes emerging before and after the band-topological transition, and analysed their wave-function profiles. This allows us to distinguish between two distinct types, the smooth and abrupt Majoranas, each with characteristic spin textures. We also showed that the nanowire spectrum is a rich fingerprint of the nanowire inhomogeneities. From the spectrum it is possible to extract information about the type of pairing and potential inhomogeneities in the nanowire. For example, a Zeeman splitting that starts with zero or finite slope at $B=0$ can distinguish between uniform and non-uniform pairing in the nanowire. Similarly, a lone Andreev level detaching into the gap as a function of $B$ reveals non-uniform and smooth electrostatic potentials. 

We have finally studied in depth the protection to local perturbations of  Majorana zero modes, and its relation to wave-function overlaps and nonlocality. As a result, we obtain several expressions for the degree of nonlocality, differing in the role of internal degrees of freedom of the spinorial wavefunction. We study their evolution with nanowire parameters and Zeeman field. The different susceptibilities $\delta N$, $\Omega_0$ and $\Omega_s$ essentially coincide for globally topological nanowires, and match the purely spatial definition $\Omega_\mathrm{max}$, but significantly differ in nanowires with non-uniform topology. 
The $\Omega$'s can be minimised in smooth NS or S'S junctions before even crossing into a topological superconductor phase. Once established, and regardless of the underlying mechanism, a small $\Omega$ protects states at zero energy, and suppresses their decoherence due to a noisy environment. 
Thus, the wave-function overlap emerges as the only relevant figure of merit of Majorana zero modes in isolated inhomogeneous nanowires.

Spatial nonlocality is intrinsically difficult to measure. The local-detection scheme proposed in Refs. \cite{Prada:PRB17,Clarke:PRB17} and analysed in detail here is much simpler than alternative schemes based on interferometry~\cite{Hell:PRB18} or {spatially correlated measurements~\cite{Liu:PRB13,Zocher:PRL13,Li:SR14,Zocher:PRL13,Haim:PRL15,Haim:PRB15, Moore:PRB18}.} 
Unlike the latter, however,  the predictive power of the local detection scheme is merely statistical. In this work we have assessed the accuracy, in a statistical sense, of local quantity $\eta$ as an estimator of the spin-uncorrelated susceptibility $\Omega_s$, as the most conservative, physically motivated measure of Majorana nonlocality. Its accuracy is rather high, particularly in the case of uniform nanowires. The significance of this for current experiments is large, as it quantifies the likelihood that a zero bias anomaly observed in transport is connected to a nonlocal Majorana zero mode. We have also analysed carefully the extent to which the estimator $\eta$ remains valid in the presence of smooth inhomogeneities. We found that for large smooth inhomogeneities with $\Delta\mu >1$ meV (of the order or greater than the superconducting gap) its accuracy is lessened, although only weakly, statistically speaking. Even if $\Delta\mu$ is very large, however, $\eta$ can still provide an upper bound for $\Omega_s$. A small $\eta\lesssim 0.2$ is a statistical guarantee that the overlap should remain bounded to $\Omega_s\lesssim 0.4$.

We have finally considered the effect of smoothness in inhomogeneous nanowires in connection to the wavefunction and spin density of Majorana zero modes. A smooth interface NS or S'S interface creates smooth Majoranas with uniform spin. These remain confined at the interface regardless of whether one of its two sides is ungapped (NS) or not (S'S). We also note that at a smooth insulating barrier, the uniform spin-polarisation of smooth Majoranas leads to their spatial separation due to a spin-dependent barrier penetration, and a suppression of their overlap as the smoothness increases. Likewise, near-perfect Andreev reflection at smooth S interfaces leads to near-equal particle and hole amplitudes, suppressed charge and a correspondingly small sensitivity to electrostatic perturbations, despite their apparently local wavefunctions. This highlights the strong connection between internal spin and particle/hole degrees of freedom, nonlocality and protection in smoothly inhomogeneous nanowires.

\acknowledgements
We acknowledge financial support from the Spanish Ministry of Economy and Competitiveness through Grant Nos. FIS2015-65706-P, FIS2015-64654-P and FIS2016-80434-P (AEI/FEDER, EU), the Ram\'on y Cajal programme, Grant Nos. RYC-2011-09345 and RYC-2013-14645 and the ``Mar\'ia de Maeztu'' Programme for Units of Excellence in Research and Development (MDM-2014-0377)

\bibliography{biblio}

\end{document}